\documentclass{elsart}
\usepackage{amssymb}
\begin{document}

\begin{frontmatter}

\title
{Phases and Density of States\\ in a Generalized Su-Schrieffer-Heeger
Model}

\author[address1,address2]{Khee-Kyun Voo},
\author[address2,address3]{Chung-Yu Mou}

\address[address1]{Texas Center for Superconductivity at the University of
Houston, Houston, TX 77204, USA
}
                                                                                
\address[address2]{Department of Physics,
National Tsing-Hua University, Hsinchu 30043,
Taiwan
}
                                                                                
\address[address3]{National Center for Theoretical Sciences,
P.O.Box 2-131, Hsinchu, Taiwan}
                                                                                
\begin{abstract}

Self-consistent solutions to a generalized Su-Schrieffer-Heeger model on
a 2-dimensional square lattice are investigated. Away
from half-filling, spatially inhomogeneous phases are found. Those phases
may have topological structures on the flux order, large unit cell bond
order, localized bipolarons, or they are simply short-range ordered and
glassy. They have an universal feature of always possessing a gap at the
Fermi level. 

\end{abstract}

\end{frontmatter}


The Su-Schrieffer-Heeger (SSH) model \cite{HKS88} was motivated by the
quasi-1-dimensional conducting polymer polyacetylene. The model exhibits
Peierls instability at the characteristic wavevector $2k_F$, where $k_F$
is the Fermi wavevector that depends on the band filling. As a result the
translational symmetry of the lattice is spontaneously broken and a gap is
opened up at the Fermi level of the electronic spectrum. Since the lattice
distortion in the ground state is multiply degenerated, topological
mid-gap soliton states can be formed at the domain boundaries. In-gap
polaron states can also be formed as soliton-antisoliton bound states. At
the empty band limit, bipolaron states are favored. The above mentioned
states as the elementary excitations of the system and their experimental
consequences were reviewed in detail in an article by Heeger $et~al.$
\cite{HKS88} In dimensions higher than one, for instance in a 2-leg
ladder, a new mid-gap soliton named the ``twiston'' was discovered
\cite{FMK01}. The ground state in the half-filled 2-dimensional (2D)
square lattice was also studied within an assumption of a small unit cell
\cite{Maz89.Maz87,TH89.TH88,YK02}, or using some more elaborated numerics
\cite{OH00}. In general, studies of the model in higher dimensions are
relatively fewer, and knowledge of the possible phases is limited.
Therefore investigations of the phases in some higher dimensional
SSH-type models are desirable. In this paper, we have extended these
studies to a generalized SSH model in two dimension (2D). It is found
that generally speaking, the types of phase are much more diverse in 2D.
Some of them appear to be nice generalizations of the above-mentioned
states in one dimension (1D). In particular, the translational symmetry is
also spontaneously broken and the Fermi levels are gapped in many phases. 
Thus this generalized SSH model represents an insightful generalization of
the conventional model and merits serious consideration.


We start by considering the conventional SSH model in the adiabatic limit, 
\begin{eqnarray}
\displaystyle H_{\rm SSH} = && - \sum_{{\bf i} \sigma} \sum_{{\bf n} = \pm
{\hat x}, \pm {\hat y} } (t+\alpha \phi_{{\bf i+n,i}})
{c_{{\bf i}\sigma}}^{\dagger} c_{{\bf i+n}, \sigma} \nonumber \\ && +
{\kappa \over 2} \sum_{\bf i} \sum_{{\bf n}={\hat x},{\hat y}} |\phi_{\bf
i,i+n}|^2,
\end{eqnarray}
where $t$ is the hopping, $\alpha$ is the electron-lattice coupling,
$\kappa$ is the interatomic elastic constant, and $\phi_{\bf ji} \equiv a
- |r_{\bf i}-r_{\bf j}|$ is the deviation of an interatomic distance from
its equilibrium value $a$. Every quantity here is real-valued and $\alpha$
and $\kappa$ are positive. To find the zero temperature configuration of
the lattice, $\phi_{{\bf i+n,i}}$'s are treated as variational parameters
and adjusted (through some procedures such as the steepest-descent method)
to minimize the total energy $\langle H_{\rm SSH} \rangle$. Alternatively,
one can recast the minimization problem into an equivalent
self-consistency problem using $\partial \langle{H_{\rm SSH}}\rangle /
\partial\phi_{\bf i,j} = 0$, which leads to $\phi_{\bf ij} = {
(2\alpha}/\kappa) \sum_\sigma < {c_{{\bf i} \sigma} }^\dagger c_{{\bf j}
\sigma}>$ and $\phi_{\bf i+n,i}$ is now a bond parameter to be
solved by numerical iteration. Defining $\Phi_{\bf i,j} = \alpha \phi_{\bf
i,j}$ and $\lambda = 4\alpha^2/\kappa$, the problem is reformulated as
\begin{eqnarray}
\displaystyle H_{\rm SSH} = && - \sum_{{\bf i} \sigma} \sum_{{\bf n} = \pm
{\hat x}, \pm {\hat y} } (t+\Phi_{{\bf i+n,i}}){c_{{\bf i}
\sigma}}^{\dagger} c_{{\bf i+n}, \sigma} \nonumber \\ && + {2 \over
\lambda} \sum_{\bf i} \sum_{{\bf n}={\hat x},{\hat y}} {|\Phi_{\bf
i,i+n}|^2}, \label{gssh1} \\ \Phi_{\bf ij} = && {\lambda \over 2}
\sum_{\sigma} <{c_{{\bf i} \sigma}}^{\dagger} c_{{\bf j} \sigma}>.
\label{gssh2}
\end{eqnarray}
In this form, it is easy to see that the only quantity characterizing the
problem is $t/\lambda$ or ${\kappa}t/(4\alpha^2)$, which is dimensionless.


In the formulation Eq.~\ref{gssh1} and \ref{gssh2}, it will be meaningful
to extend $\Phi_{\bf ij}$ from the $real$ number regime into the $complex$
number regime. The physical appeal of the extension is that the model now
addresses a class of more general problems by the following reason.  
Generally, the parameter $\Phi_{\bf ij}$ represents an overlap integral
$<\psi_{\bf i}|H'|\psi_{\bf j}>$ describing off-site electronic
correlations, with $H'$ and $|\psi_{\bf i/j}>$ being the off-site part of
the total Hamiltonian and the local atomic wavefunctions respectively.
Since in addition to the electron-lattice interaction, a renormalized
low-energy effective theory may contain other many-particle effects
manifested in the overlap integral, there is no reason for $\Phi_{\bf
ij}$ to be necessarily real when these effects are included. This
extension thus goes beyond the electron-lattice interaction and it may be
viewed as a prototype of this class of problems. Indeed, a four-fermion
interaction is generated at integrating out $\Phi_{\bf ij}$, and therefore
the self-consistency equation that $\Phi_{\bf ij}$ respects can also be
regarded as a mean field equation. In fact, the model is formally the same
as a mean field theory in the high-$T_{\rm c}$ cuprates \cite{AM88.MA89}.
In view of this connection, we have in this paper investigated this
generalized SSH model in detail. On one hand, our study is a
generalization of the conventional SSH model in 2D; on the other hand, it
is also a study complementary to those mean field studies of the
high-$T_{\rm c}$ system, where a different parameter regime (larger
$t/\lambda$) was emphasized.


\begin{figure}
\includegraphics{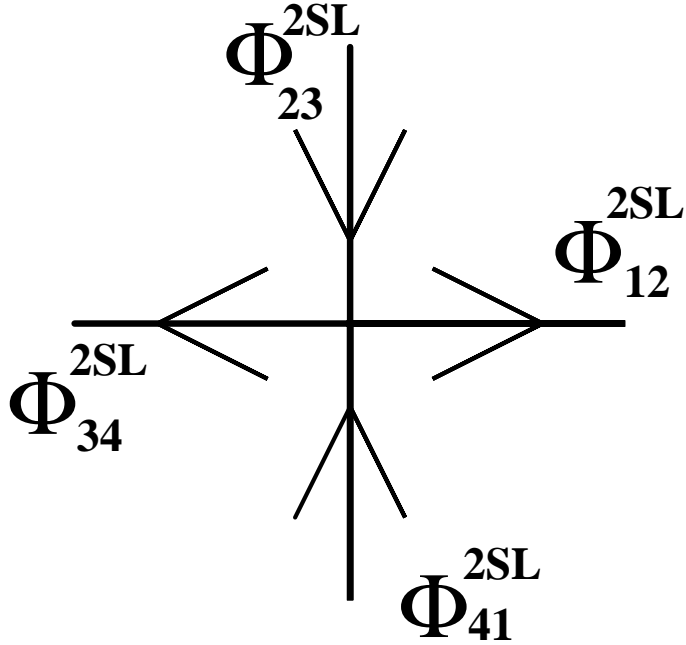}
\vspace{300pt}
\caption{The 2-sublattice ansatz. The 2D square lattice is patched up by
centering this cross at sites belong to one of the two sublattices. The
complex-valued nature is indicated by an arrow, which means that 
${\Phi_{ij}^{\rm 2SL}}^{\dagger} = \Phi_{ji}^{\rm 2SL}$ is the reversed
arrow and may not be identical to $\Phi_{ij}^{\rm 2SL}$.} 
\label{fig00}
\end{figure}

It is noted that the generalized hamiltonian bears the same form as a mean
field hamiltonian in the high-$T_{\rm c}$ cuprates \cite{AM88.MA89}, and
presumably they should bear the same solution. In the high-$T_{\rm c}$
context, a 2-sublattice (2SL) ansatz (see Fig.~\ref{fig00}) designed
to capture any $(\pi,\pi)$-instability was adopted in
Ref.~\cite{AM88.MA89} for the solution. Such a high translational
symmetry solution might be appropriate in that context, since other
effects such as fluctuation might at work and preserve the symmetry to a
high degree, but in our case, where we expect soliton or polaron
solutions, such an ansatz at the outset is clearly inappropriate.
Nevertheless, we still have reproduced the solution within that framework
and compare with the result of our approach which allows spatially
nonhomogeneous solutions. The 2SL ansatz assumes the distribution of
$\Phi$'s on the lattice has a repeating unit as shown in Fig.~\ref{fig00}.
Different types of solution were found \cite{AM88.MA89}, they were the
staggered-Peierls (SP) with complex $\Phi^{\rm 2SL}$'s and $\Phi_{12}^{\rm
2SL} = \Phi_{23}^{\rm 2SL} = \Phi_{34}^{\rm 2SL}$ but $|\Phi_{41}^{\rm
2SL}| > |\Phi_{34}^{\rm 2SL}|$; uniform staggered-flux (u-SF) with all the
$\Phi^{\rm 2SL}$'s are equal and have imaginary components; uniform
real-bond (u-R) with all the $\Phi^{\rm 2SL}$'s are equal and real; kite
(K) with all the $\Phi^{\rm 2SL}$'s are real and either $\Phi_{12}^{\rm
2SL} = \Phi_{23}^{\rm 2SL} \neq \Phi_{34}^{\rm 2SL} = \Phi_{41}^{\rm 2SL}$
or $\Phi_{12}^{\rm 2SL} = \Phi_{34}^{\rm 2SL} \neq \Phi_{23}^{\rm 2SL} =
\Phi_{41}^{\rm 2SL}$.


We have found the zero temperature self-consistent solutions
iteratively on periodic boundary finite square lattices. Each bond is
treated as an independent parameter. Meanwhile, phases and energies
within the 2SL ansatz are also solved and compared with the result of
our unrestricted search. Our unrestricted-bond iteration converges to the
2SL form only at two regimes, (i) $x \sim 0$ and $t > t_{\rm c} \sim
0.1\lambda$, where the uniform-SF is located at; and (ii) large $t$ and
$x$, where the uniform real-bond phase is located at. $x$ is the hole
concentration between 0 and 1. 

\begin{figure}
\includegraphics{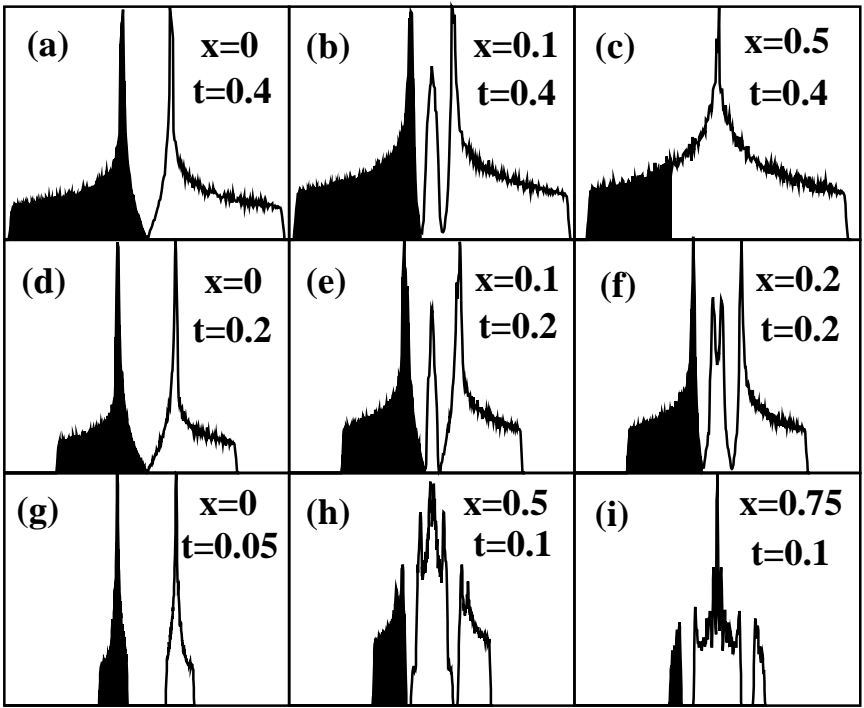}
\vspace{300pt}
\caption{Total density of states at some representative dopings and
hoppings ($\lambda \equiv 1$): The shaded are occupied levels and the
horizontal
axis of each box is the energy from $-2.5$ to $+2.5$.}
\label{fig01}
\end{figure}

Away from the above regimes, our solutions generally do not fit in a 2SL
ansatz. Some representative phases which we will discuss and their
energies (per site) $E_G$ are grouped here beforehand. In the brackets,
energies $E_G^{2SL}$ and phases from the 2SL ansatz are also given for
comparison. We obtain dimer-box (D-B) glass phase at, $t=0$, $x=0.3$,
with $E_G=-0.1750\lambda$ [$E_G^{2SL}=-0.1608\lambda$, K]; $t=0$, $x=0.5$,
with $E_G=-0.1250\lambda$ [$E_G^{2SL}=-0.1077\lambda$, u-R]. We find
striped-staggered-flux (s-SF) at, $t/\lambda=0.2, x=0.1$, with
$E_G=-0.5017\lambda$ [$E_G^{2SL}=-0.4991\lambda$, u-SF]; $t/\lambda=0.2,
x=0.2$, with $E_G=-0.4784\lambda$ [$E_G^{2SL}=-0.4733\lambda$, u-R];
$t/\lambda=0.4, x=0.1$, with $E_G=-0.8134\lambda$ 
[$E_G^{2SL}=-0.8088\lambda$, u-SF]. Wigner lattice of bipolarons
(WBP) is found at, $t/\lambda=0.1, x=0.75$, with $E_G=-0.1267\lambda$
[$E_G^{2SL}=-0.1232\lambda$, u-R]. A bond order wave (BOW) which we have
named Real-staggered-box (RSB) is found at, $t/\lambda=0.1, x=0.5$, with
$E_G=-0.2402\lambda$ [$E_G^{2SL}=-0.2390\lambda$, u-R]; $t/\lambda=0.05,
x=0.5$ with $E_G=-0.1798\lambda$ [$E_G^{2SL}=-0.1734\lambda$, u-R].
Solutions of lower translational symmetry, when they exist, are found to
have lower energies and the percentage of difference is more prominent at
small $t$ but intermediate $x$. Typical total density of states (DOS) of
these phases are shown in Fig.~\ref{fig01}.


$Dimer$-$box$ $glass$: On the $t=0$ axis, a solution of high degeneracy
can be found analytically. It is straightforward to verify that a lattice
$arbitrarily$ filled with disjointed dimers and boxes is a solution
\cite{Sac90}. A dimer \cite{AM88.MA89,DK89} is a doubly filled
nearest-neighbor bond \cite{VM02.7} with $|\Phi|=\lambda/2$. A box
\cite{VM02.1} is a plaquette with bonds 
\begin{eqnarray}
|\Phi_{12}|=|\Phi_{34}|,~~~~ |\Phi_{23}|=|\Phi_{41}|,\\ 
\displaystyle |\Phi_{12}|^2+|\Phi_{23}|^2 = ({\lambda \over 2})^2,
\label{db}
\end{eqnarray}
and a real 
\begin{eqnarray}
\Phi_{12}\Phi_{23}\Phi_{34}\Phi_{41} < 0, 
\end{eqnarray}
where 1, 2, 3, and 4 are consecutive corners of the plaquette. The lowest
two levels of a box are filled with 4 electrons. Both dimer and box have
an average of 1 electron per site, energy $-\lambda/4$ per site, and only
two levels at $\pm \lambda/2$. At $x=0$, the lattice is fully filled with
dimers and boxes. At $x>0$, a solution can be simply obtained by plucking
off dimers or boxes \cite{VM02.2} from the lattice and the energy will be
just $-(1-x)\lambda/4$. Creating empty sites also creates zero energy
states, therefore the spectrum of a doped zero-$t$ system is a three-level
structured DOS $D (\varepsilon) = x \delta(\varepsilon) + (1-x)
[\delta(\varepsilon+\lambda/2) + \delta(\varepsilon-\lambda/2)]$. Since
the arrangement of the dimers and boxes is arbitrary, this phase is in
general a glass of bond and plaquette centered electron charges. But it is
important to note that due to the degeneracy condition Eq.~\ref{db}, a
box can be continuously deformed into two dimers or vice versa without any
energy barrier intervening. Therefore this glass might not be a stable
phase when fluctuation effect is taken into account, and some 
translational symmetry is expected to be restored.

It can be shown analytically that the D-B glass has a lower energy than
the u-R phase at $t=0$ and $x \rightarrow 1$. At $x \rightarrow 1$, the
kinetic energy of a u-R phase is simply obtained by filling in the band
bottom, i.e., $-2D(t+\Phi)(1-x)$, and the potential energy is
$2D\Phi^2/\lambda$, where $D=2$ is the dimensionality. Minimizing the
total energy gives $\Phi=(1-x)\lambda/2$, hence energy $E_G^{2SL} =
-2tD(1-x) - ({\lambda}D/2)(1-x)^2$. It is of ${\mathcal{O}}[(1-x)^2]$ at
$t=0$, while the D-B glass has $E_G=-(1-x)\lambda/4$ of
${\mathcal{O}}[(1-x)]$. Comparing $E_G^{2SL}$ and $E_G$ also gives a scale
$t=\lambda/(8D)$ at which the D-B glass is destabilized and there is a
crossover from short-range-order (SRO) to long-range-order (LRO).

At intermediate $x$, one can only obtain $E_G^{2SL}$ numerically. We have
studied, e.g., $x=0.1, 0.3$, and $0.5$, and have found that they have
energies (which are given before, and are stable at lattices 40$\times$40,
80$\times$80, and 120$\times$120) substantially higher than the D-B glass.  
A feature to note is that the 2SL phases are ungapped, whereas the
three-level D-B glass is obviously gapped. The levels expand into bands at
small but nonzero $t$ and the gapping remains, as a vestige of the
zero-$t$ phase.


\begin{figure}
\includegraphics{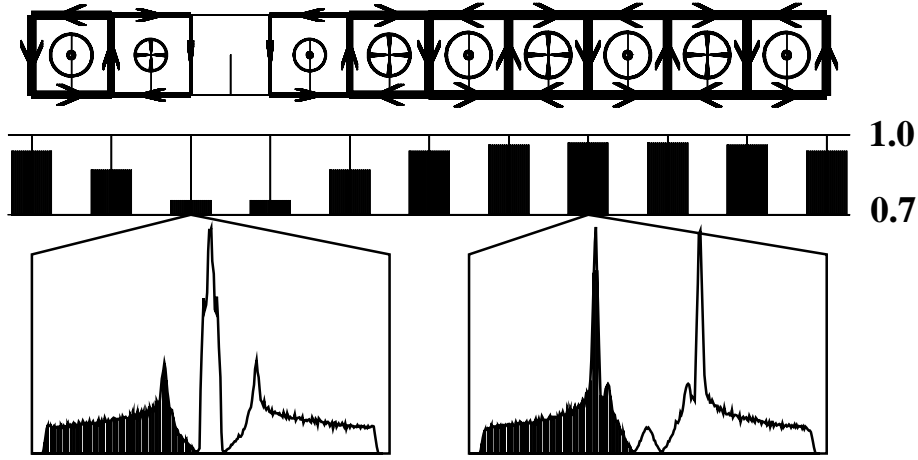}
\vspace{250pt}
\caption[*]{Topological stripe: (Top) The bond current
$(4t/\lambda)$Im$\Phi_{\bf ij}$, orbital moment (average of directed bond
currents round a plaquette), and (middle) electron density profiles across
an 1D antiphase domain wall ($t/\lambda=0.2, x=0.1$). Bond
thicknesses are proportional to the current sizes and circle areas to the
moment sizes. Maximum bond current here is 0.106/s, and maximum
moment is 0.105/s. The domain wall shown here contains neutral
plaquettes. (Bottom) Typical local DOS at hole-rich and hole-poor sites
are also shown.}
\label{fig02}
\end{figure}

$Striped$-$staggered$-$flux$: At the regime $0 \lesssim x \lesssim 0.2$
and $t_{\rm c} (\sim 0.1\lambda) < t \lesssim 0.6\lambda$, the holes are
found to arrange into stripes along the lattice axes \cite{VM02.3,VM02.4}
(say $y$-axis, see Fig.~\ref{fig02}). Whereas uniform-SF is obtained
within the 2SL ansatz. The stripes at the meantime are also topological
domain walls separating antiphase SF domains \cite{VM02.4}. Two types of
domain walls are found, one has a column of neutral-flux plaquettes
inserted, the other has a column of plaquettes deleted. It is remarkable
that the electron depletion can be as large as $\sim$0.2 and has a short
coherence length of only a few lattice constants. We also make an
observation that it has one hole per unit length \cite{VM02.6}. The
spectrum [see Fig.~\ref{fig01}(b), (e), and (f)] is nodally gapped at the
Fermi level and the gap maxima at $(0,\pm \pi)$ and $(\pm \pi,0)$
\cite{AM88.MA89,DK89,HMA91,UL92,CLM01,SD02} manifest themselves as the
van Hove singularity peaks in the DOS. Local DOS shows that the mid-band
has a great part composed of states localized along the stripes (see
Fig.~\ref{fig02}).

It is interesting to check if such striped-SF phase have energies lower
than that of the 2SL phases. For some commensurate fillings like $x=0.1,
0.2$, and 0.3, we have obtained straight hole-stripes along one of the
lattice axes. In such cases, the fact that the striped-SF phase have
lower energy can be justified in the thermodynamic limit as follows.
Owing to the translational invariance along one axis, we may
equivalently seek for such solution within an ansatz with translational
symmetry $\Phi_{{\bf i},{\bf i}+{\hat \alpha}}=\Phi_{{\bf i}+2{\hat
y},{\bf i}+{\hat \alpha}+2{\hat y}}, {\hat \alpha}={\hat x},{\hat y}$
\cite{VM02.5}. Due to the reduction of numerical effort (diagonalizing a
hamiltonian on a $N_x$$\times$$N_y$ lattice is now separately
diagonalizing $N_y/2$ hamiltonians on $N_x$$\times$$2$ lattices),
self-consistent solutions can now be pursued on much larger lattices. We
have justified that the striped-SF at, e.g., $t/\lambda=0.2, x=0.1, 0.2$,
and $t/\lambda=0.4, x=0.1$ have energies (which are given before) lower
than that of the 2SL solutions. The energies are stable at lattices from
40$\times$40 to 120$\times$120.

Note that the stripes here are not a result of frustrated phase
separation since we assume no repulsion between holes. They are more
appropriately understood as soliton states created to accommodate the
holes, analogous to the soliton states in the 1D SSH model \cite{HKS88}.
By segregating out the holes, the SF order has survived into higher hole
concentrations where it would not has existed within the 2SL ansatz.


$LRO$-$SRO$ $crossover$: The nonzero but small $t$ regime, $t < t_{\rm c}
\sim 0.1\lambda$ is a crossover region between the large-$t$ regime of LRO
and the zero-$t$ regime of degenerated glass. SRO dominates over this
regime and there are lots of almost-degenerate local minimum solution.
Phases in
this regime are generally glassy, have complex bonds at $x<0.3$, and real
bonds at $x>0.3$. Charge distribution is correlated with the local bond
order and also nonuniform in general. Basically, regions that are crowded
with electrons may have complex bonds and regions that are not, have
essentially real bonds. In our calculation, such glassy states found by
iteration do have energies lower than that of the 2SL solution and their
spectra are always gapped at the Fermi level.

\begin{figure}
\includegraphics{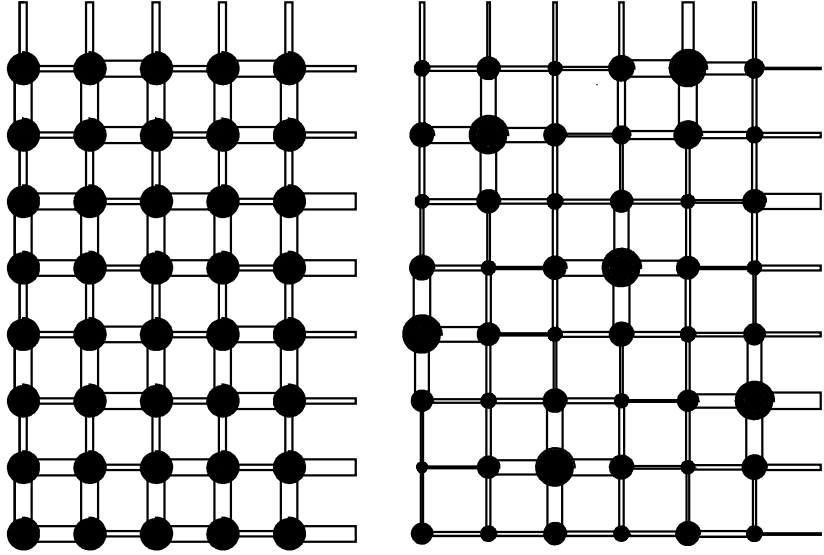}
\vspace{250pt}
\caption[*]{(Left) Real-staggered-box BOW with uniform charge distribution
at $t/\lambda=0.1, x=0.5$. (Right) Wigner lattice of bipolarons at
$t/\lambda=0.1, x=0.75$. Bond thicknesses and circle areas are drawn
proportional to the bond magnitudes and electron densities respectively.}
\label{fig03}
\end{figure}

At $x \sim 1$ of this crossover regime, the obtained state contains
uncorrelatedly scattered bipolarons. A bipolaron is a localized state
occupied by two electrons, with nonzero real bonds at its vicinity.
Decreasing $t$ shrinks them into dimers, or increasing electron density
turns them into an almost-hexagonal closed-packed Wigner lattice (WBP)
(see Fig.~\ref{fig03}). In the well-isolated bipolarons regime, the energy
can be readily obtained as the sum of the energies of the individual
bipolarons. For the Wigner lattice, for instance at $t/\lambda=0.1,
x=0.75$, we obtain the same energy (which is given before) at lattices
from 20$\times$20 to 32$\times$32, and it is lower than the energy of the
2SL solution (also given before).

Two exceptions out of these glassy phases are the BOWs with uniform
charge distribution named ``real-staggered-box'' (RSB) (see
Fig.~\ref{fig03}) found at $x=0.5$, and ``box-staggered-flux'' (B-SF) at
$x=0$. RSB found at $t/\lambda=0.1$ and $0.05$ have energies (which are
given before, and are stable at lattices from 40$\times$40 to
120$\times$120) lower than that of the u-R. B-SF is a phase with SF order
and stronger bonds distributed like a square array of boxes (resembles
those boxes in Ref.~\cite{AM88.MA89}), and it also has energies lower than
that of SP. 

When $x < 0.3$, bonds are complex but remain glassy. The spectrum has a
narrow in-gap band consists of almost-localized states. These in-gap
states are fundamentally different from those of the striped-SF at larger
$t$. The latter are soliton states supported by the LRO, while the former
are states localized in the bond glass. 


\begin{figure}
\includegraphics{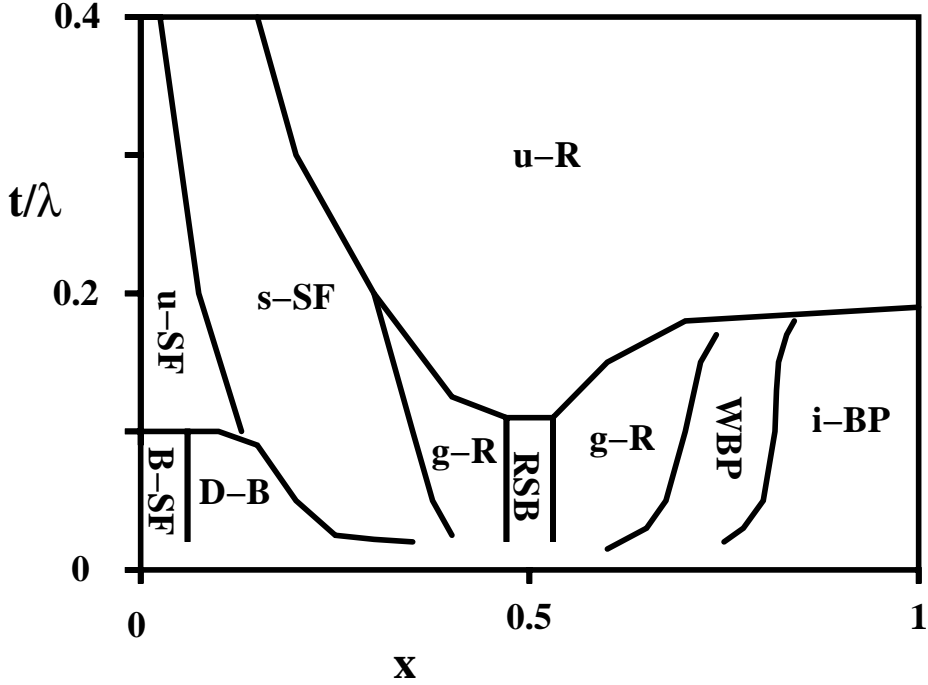}
\vspace{300pt}
\caption[*]{Approximate phase diagram:
B-SF: {\bf B}ox-{\bf s}taggered-{\bf f}lux (gapped).
u-SF: {\bf U}niform {\bf s}taggered-{\bf f}lux (ungapped when $x>0$).
s-SF: {\bf S}triped-{\bf s}taggered-{\bf f}lux (gapped).
u-R: {\bf U}niform {\bf r}eal-bond (ungapped).
D-B: {\bf D}imer-{\bf B}ox glass (gapped). The dimer and box order are
exact only at $t=0$.
g-R: Phases with {\bf g}lassy {\bf r}eal-bond and charge density
(gapped), and are not characterizable by a simple order.
RSB: {\bf R}eal-{\bf s}taggered-{\bf b}ox BOW with uniform charge density
(gapped).
WBP: {\bf W}igner lattice of {\bf b}i{\bf p}olarons (gapped).
i-BP: {\bf I}solated and uncorrelated {\bf b}i{\bf p}olarons (gapped).}
\label{fig04}
\end{figure}

In summary, an approximate phase diagram is given in Fig.~\ref{fig04}.
Spatially nonhomogeneous phases are shown to be energetically more
favorable in the generalized SSH model in 2D away from half-filling, and
the electronic spectrum is believed to show a similar gapped Fermi level
as that in the
1D conventional SSH model. The doped holes also go into the in-gap states
and topological object may also be formed, but the 2D phases are much more
complicated than the simple polymerizations in 1D. Those phases may be
superstructured or short-range-ordered. At large $t/\lambda$ and large
$x$, the translational symmetry is preserved and this is fundamentally
different from the 1D case, where the Peierls instability always in
effect.


We thank the support from the National Science Council of Taiwan under
grant no. NSC91-2112-M-007-049, and National Center for Theoretical 
Sciences (Phys. Div.) of Taiwan for letting us to use their facilities.
KKV also thanks the support from the Texas Center for Superconductivity at
the University of Houston and a grant from the Robert Welch Foundation.

%
%



\begin{thebibliography}{99}
\bibitem{VM02.1} We note that the box order discussed here is a
more general case of that discussed in Ref.\cite {AM88.MA89} and \cite
{DK89} which has $|\Phi_{12}|=|\Phi_{23}|=|\Phi_{34}|=|\Phi_{41}|$.
\bibitem{VM02.2} If the number of electron is odd, it is energetically
more favorable for the unpaired electron to form a singly filled dimer
with $|\Phi|=\lambda/4$ and energy $-\lambda/8$, rather than breaking the
bond entirely forming an empty site and a nonbonded electron with energy
0. In the former case the singly filled dimer contains charge 1 and spin
1/2 (``polaron''). In the later case the empty site contains charge
1 (taking the half-filled site as reference) but spin 0 (``holon''), and
the unpaired electron contains spin 1/2 but charge 0 (``spinon'').
\bibitem{VM02.4} If the hole number in the finite lattices does not
support hole-strings long enough to span the periodic lattice, they form
smaller closed hole-strings and are still antiphase boundaries of the SF
domains.
\bibitem{VM02.7} The dimers are in fact bipolarons of the smallest size.
\end{thebibliography}
\end{document}